# Localised IR spectroscopy of hemoglobin

Fiona Yarrow and James H. Rice[a]

NanoPhotonics Research Group, School of Physics, University College Dublin, Belfield, Dublin, Ireland

a) Electronic mail: james.rice@ucd.ie

**Abstract**

IR absorption spectroscopy of hemoglobin was performed using an IR optical parametric oscillator laser and a commercial atomic force microscope in a novel experimental arrangement based on the use of a bottom-up excitation alignment. This experimental approach enables detection of protein samples with a resolution that is much higher than that of standard IR spectroscopy. Presented here are AFM based IR absorption spectra of micron sized hemoglobin features.

**Introduction**

Absorption spectroscopy is a widely applied technique for chemical characterisation. This method is able to detect both luminescent and non-luminescent materials and provide chemical specific information. An extensively used form of absorption spectroscopy is infrared absorption (IR) spectroscopy. This measures specific frequencies in the infrared region of the electromagnetic spectrum at which constituent parts of molecules corresponding to specific types of molecular bonds vibrate. This makes possible structural elucidation and compound identification of materials. As a consequence, IR absorption is extensively used as an analytical tool.

IR spectroscopy has been widely applied for the molecular structure characterization of lipids and proteins. As outlined IR spectroscopy detects molecular vibrations accompanied by changing molecular dipole moments. As a consequence the vibration frequencies that are detected are sensitive to molecular conformation. Thus, spectroscopy is a useful method to investigate lipid structure and conformation.

While IR absorption studies of biosystems have been well established in the study of proteins there remains a number of limitations in the information that IR absorption spectroscopy can provide. For example, when studying mircon or nanosized features IR absorption spectroscopy is limited by the diffraction limit (Abbe 1873; Rice 2007). The maximum image resolution in optical microscopy is found to be ca. $\lambda/2$. As a result of diffraction, the image resolution will be 2.5 μm when imaging in the infrared using electromagnetic radiation at 5 μm (corresponding to an IR absorption frequency of 2000 $cm^{-1}$). This means that IR absorption spectroscopy technology cannot be applied to study features that are approximately smaller than a few microns.

A newly emerging method for IR absorption spectroscopy using a photothermal based methodology enables IR spectroscopy of smaller amounts of materials than is currently possible using established IR absorption methods. This new method uses an Atomic Force Microscope (AFM) cantilever tip as the detection mechanism

(Hammiche et al 2004; Hammiche et al 1999; Hammiche et al 2004). One particular method based on this approach, referred to as AFMIR, measures IR absorption directly via measuring local transient deformation in the sample via the AFM cantilever which is induced by an IR pulsed laser tuned at a vibration absorbing wavelength (Dazzi et al 2005; Houel et al 2007; Mayet et al 2008; Hill et al 2009; Rice 2010). This enables IR absorption spectra to be recorded of features as small as the AFM cantilever tip. Studies using this approach have reported the study of quantum dot nanomaterials with a spatial resolution of 60 nm (Houel et al 2007). To date the experimental methodology of AFMIR has utilised attenuated total internal reflection arrangement in combination with IR cyclotron radiation or a top down configuration using a customised IR laser source.

Here, we outline work performed on an experimental method for IR surface spectroscopy that samples directly via a novel bottom-up optical arrangement using a commercial laser system. The advantage of this set-up is that is uses a commercial laser rather than synchrotron radiation and allows the use of commonly used substrates such as glass or mica. In this letter, this novel AFMIR set-up is applied for the first time to study hemoglobin. The work presented here demonstrates that AFMIR can be applied to study micron sized aggregations of protein.

**Methods**

The AFMIR experimental set-up is shown in Fig 1a. The experimental configuration consisting of an optical parametric oscillator (OPO) laser and an AFM. The excitation IR radiation is directed upward in a novel configuration using gold coated mirrors to direct the laser light. The sample was mounted onto a glass side to facilitate this optical arrangement.

IR radiation was generated using an OPO laser (Cohesion) based on a periodically poled $LiNbO_3$ crystal emitting IR laser radiation that is tuneable over > 3.0 to 3.6 µm. The output power was c.a. 2 mW. The laser was focused to a relatively large spot of c.a. 500 µm on the sample in order to cover the entire area probed. The energy was low enough to avoid damaging the sample. An AFM (Veeco Explorer system) was used with a scanner with lateral and vertical dimensions of 100 x 100µm and 8µm respectively. The AFM is operated in contact mode enabling simultaneous IR and topography measurements. Silicon nitride tips mounted on a V-shaped cantilever with a nominal spring constant of 0.05 N/m (Veeco) were used. A force setpoint of 1 - 3 nN was used. Samples were prepared on standard microscope glass slides. A Fourier Transform IR (FTIR) spectrometer (Varian model 3100) was used to record a reference IR spectrum.

**Results**

The AFM tip was positioned over the sample with the tip in contact with the sample surface. Following absorption of the incident radiation, the energy absorbed is dissipated through thermal and acoustic mechanisms. Propagating acoustic waves create a deformation in the surface topography which can be detected by displacement of the AFM tip (Dazzi et al 2005; Rice 2010). As an IR laser source is tuned into resonance with a vibration mode, absorption of IR radiation increases. The response of the cantilever tip was monitored following the application of the IR radiation.

AFMIR studies of a deposited layer of hemoglobin on a glass slide were undertaken. An AFM topography image of the surface was recorded (as shown in Fig 1b). A small area of the sample surveyed in the AFM topography image (marked α in Fig 1b) was selected for study. The hemoglobin feature in this area was c.a. 1 µm higher than the surrounding layer and has a diameter of c.a. 200 nm. The lateral size of the sample area probed is proportional to the size of the AFM tip (i.e. around 20 nm).

Fig 2a shows the oscillation of the AFM cantilever following absorption of the IR laser pulse by the sample. The intensity of the cantilever oscillation changes on resonance (2960 cm$^{-1}$) and off resonance (2810 cm$^{-1}$) with the C-H stretching mode of hemoglobin. The intensity of the oscillation of the cantilever oscillation as a function of wavelength was recorded. The resulting AFMIR spectrum of hemoglobin is shown in Fig 2b.

The AFMIR spectrum is shown alongside the FTIR spectrum of hemoglobin. The two spectra show very similar features. The AFMIR spectra shown in Fig 2.b possess a wavelength resolution of 15 cm$^{-1}$, while the FTIR based spectrum has a wavelength resolution of 2 cm$^{-1}$. Comparing the AFMIR and FTIR spectra shows that they possess similar spectral features. Both spectra show the presence of peaks (marked α, β, χ, δ, ε, φ) on a broad background.

The position of these peaks corresponds to crystalline hemoglobin. Kuenstner et al (2000) reported the position of peaks for crystalline hemoglobin to be 2871.5, 2960.2 and 3060.5 cm$^{-1}$ which match the position of the peaks seen in both the AFMIR and FTIR spectra. The bands seen in the spectra arise from N-H and C-H vibrations. The amide B band at 3061 cm$^{-1}$ is assigned to an intramolecular hydrogen-bonded N-H stretching or to an overtone band (i.e. 2 x 1541 cm$^{-1}$) (Kuenstner et al 2000). The band at 2960 cm$^{-1}$ is due to aliphatic C-H stretching. Changes in the relative intensities in some peaks are present when comparing the AFMIR and FTIR spectra. Normalising to the peak at δ the peak at χ is reduced in intensity, while peak at β is reduced compared to α. These changes in relative intensity may be associated with the microsized particle measured inducing small changes in the conformation of the protein.

Studies of smaller nanosized protein layer features were undertaken. Fig 3a shows an AFM image of hemoglobin sample on a glass surface. Two regions were probed which were c.a. 150 nm different in height (denoted by α and β. Studies of these two regions showed that difference in the AFM tip oscillations on (region β) and off (region α) the heme ledge could be seen at a wavelength of 2960 cm$^{-1}$, which corresponds to on resonance (see Fig 3b). The thinner part of the sample showed a weaker AFM tip oscillation than the thicker part of the sample (i.e. regions denoted by α and β̃ This shows that small nanosized features can be discerned using AFMIR. Changing the wavelength to a wavelength that corresponds to off resonance (i.e. 2810 cm$^{-1}$) led to the same effects in the oscillations as shown in Fig 2a.

In conclusion, IR absorption spectroscopy of hemoglobin was performed using an IR optical parametric oscillator laser and a commercial atomic force microscope in a novel configuration. This experimental approach enables detection of micron sized protein features. This resolution cannot be achieved by standard IR spectroscopy. This methodology for IR spectroscopy can potentially be applied to study other protein structures on the micro- and nanometre length scales.

**Acknowledgements**

The authors would like to acknowledge Science Foundation Ireland for supporting this research.

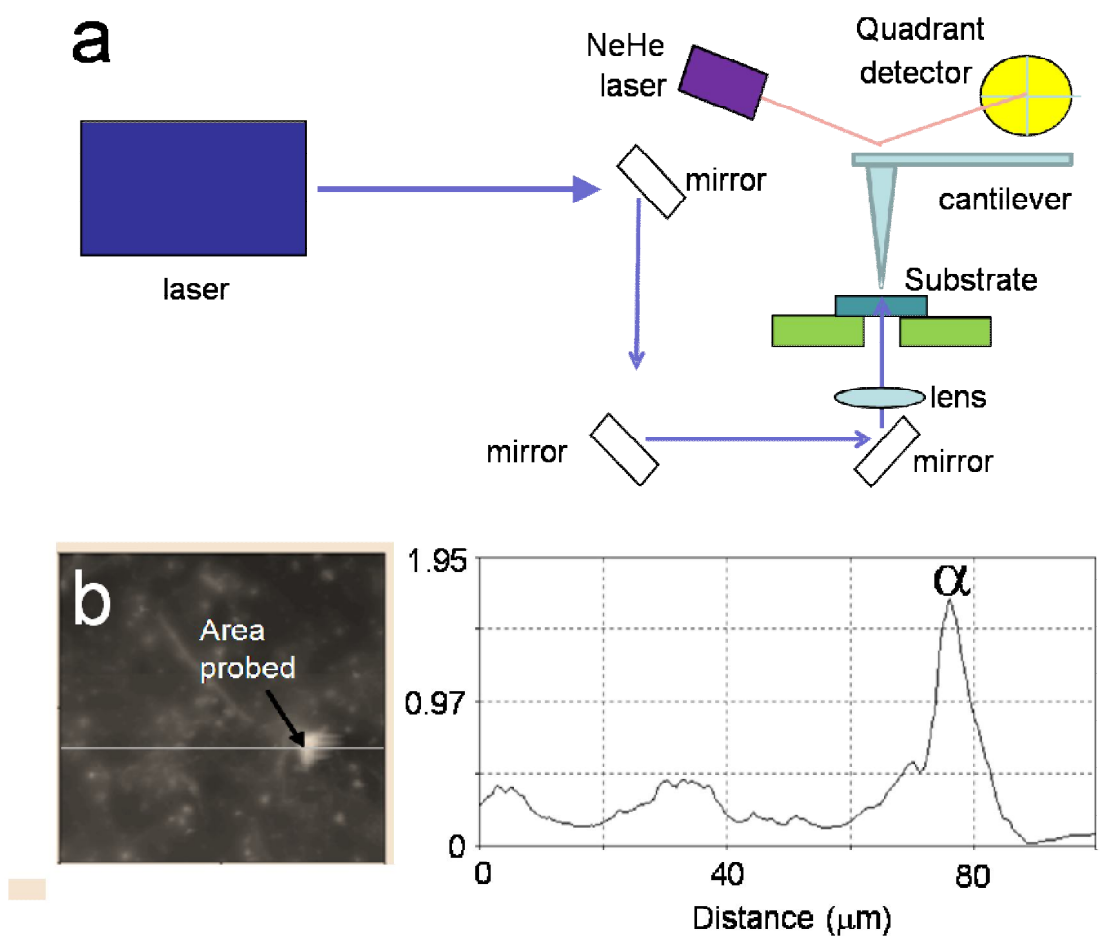

*Fig 1. a) Schematic drawing of the experimental set-up, b) AFM image of hemoglobin sample on a* glass surface (100 x 100 μm area). Left, topography and right, height profile recorded at the position of the white line. The arrow shows the region probed by AFMIR, which corresponds to α.

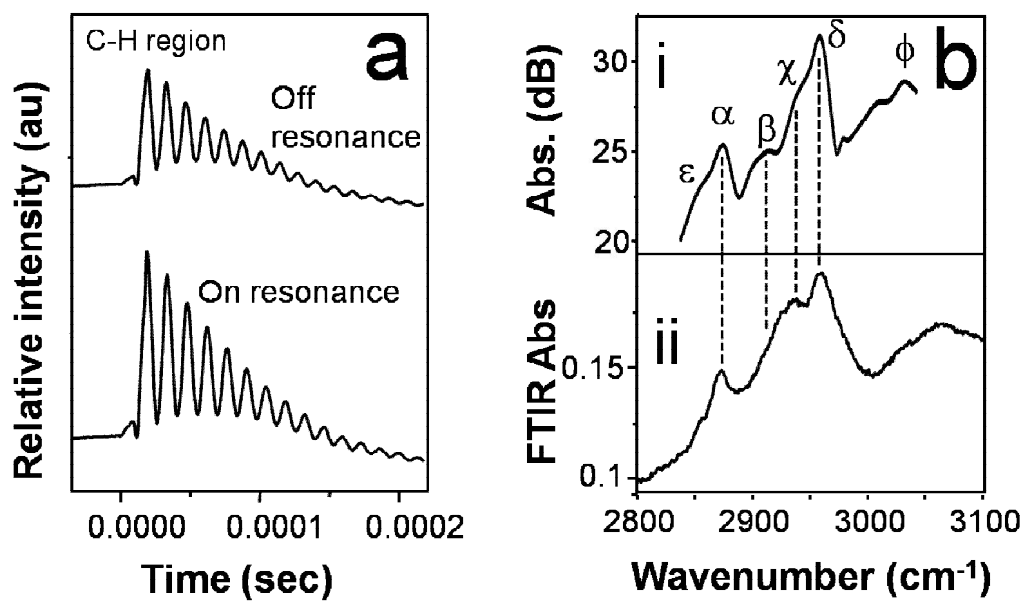

Fig 2. a) Plot of AFM tip oscillations on (2960 cm$^{-1}$) and off (2810 cm$^{-1}$) resonance, b) IR absorption spectra for hemoglobin.

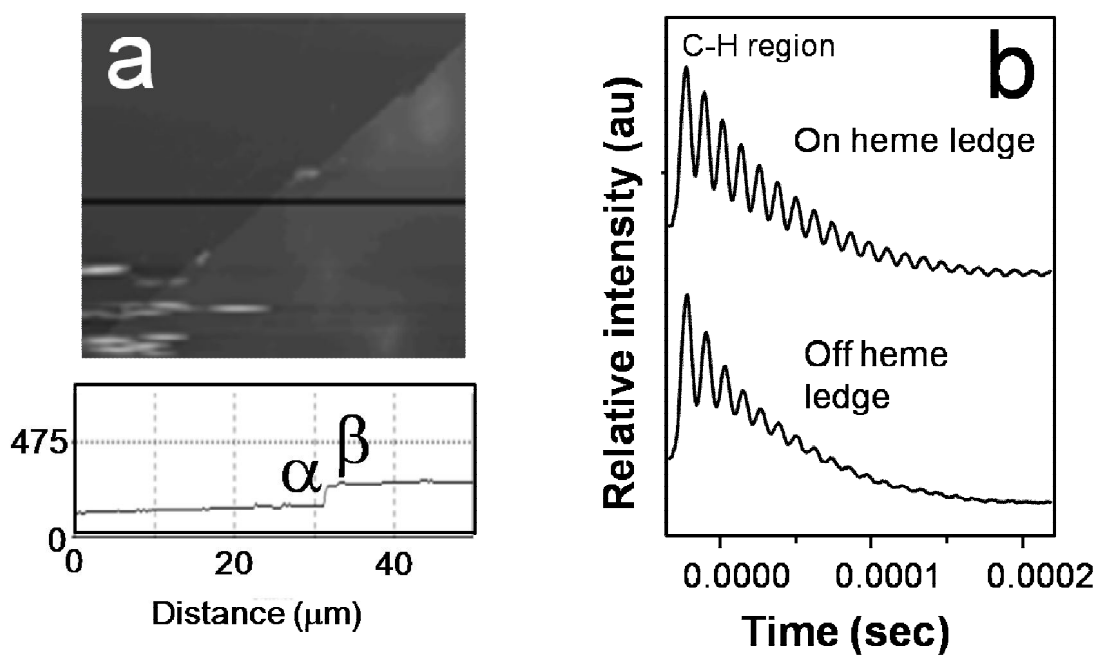

Fig 3.a) AFM image of hemoglobin sample on a glass surface (500 x 500 nm area). Top, topography and bottom, height profile recorded at the position of the black line. The areas probed are denoted by α and β. b) Plot of AFM tip oscillations on resonance (2960 cm$^{-1}$) on and off the heme ledge (β and α, respectively).